\documentclass{article}
\usepackage{ijcai24}

\usepackage{times}
\usepackage{soul}
\usepackage{url}
\usepackage[hidelinks]{hyperref}
\usepackage[utf8]{inputenc}
\usepackage[small]{caption}
\usepackage{graphicx}
\usepackage{amsmath}
\usepackage{amsthm}
\usepackage{booktabs}
\usepackage{algorithm}
\usepackage{algorithmic}
\usepackage[switch]{lineno}
\usepackage{multirow}
\usepackage{tabularx}

\title{UniRQR: A Unified Model for Retrieval Decision, Query, and Response Generation in Internet-Based Knowledge Dialogue Systems}

\author{
Zhongtian Hu$^1$
\and
Yangqi Chen$^1$\and
Meng Zhao$^1$\and
Ronghan Li$^2$\And
Lifang Wang$^1$
\affiliations
$^1$Northwestern Polytechnical University,China\\
$^2$Xidian University,China
\emails
ahxchzt@mail.nwpu.edu.cn
}

\begin{document}

\maketitle

\begin{abstract}
Knowledge-based dialogue systems with internet retrieval have recently attracted considerable attention from researchers. The dialogue systems overcome a major limitation of traditional knowledge dialogue systems, where the timeliness of knowledge cannot be assured, hence providing greater practical application value. Knowledge-based dialogue systems with internet retrieval can be typically segmented into three tasks: \emph{Retrieval Decision}, \emph{Query Generation}, and \emph{Response Generation}. However, many of studies assumed that all conversations require external knowledge to continue, neglecting the critical step of determining when retrieval is necessary. This assumption often leads to an over-dependence on external knowledge, even when it may not be required. Our work addresses this oversight by employing a single unified model facilitated by prompt and multi-task learning approaches. This model not only decides whether retrieval is necessary but also generates retrieval queries and responses. By integrating these functions, our system leverages the full potential of pre-trained models and reduces the complexity and costs associated with deploying multiple models. We conducted extensive experiments to investigate the mutual enhancement among the three tasks in our system. What is more, the experiment results on the Wizint and Dusinc datasets not only demonstrate that our unified model surpasses the baseline performance for individual tasks, but also reveal that it achieves comparable results when contrasted with SOTA systems that deploy separate, specialized models for each task.
\end{abstract}
\begin{figure}[!t]
	\includegraphics[width=1\columnwidth]{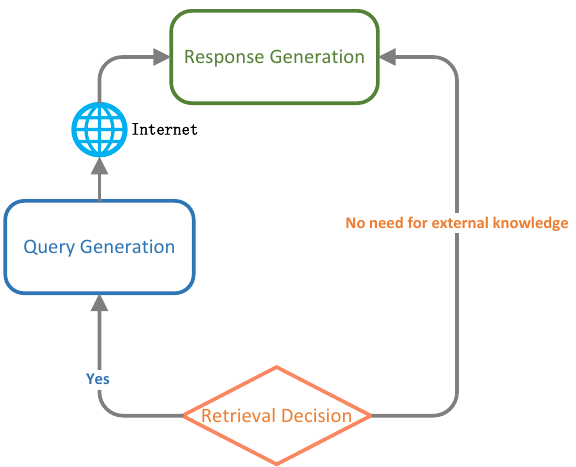}
	\caption{Workflow for Internet-based knowledge dialogue system.}
	\label{fig1}
\end{figure}
\section{Introduction}
As open-domain dialogue systems evolve \cite{ide2021multi,LIU2021106}, knowledge-based dialogue systems \cite{li2022knowledge,zhou2020kdconv,tang2023enhancing,2019Knowledge,2021Towards} are achieving remarkable progress. Unlike conventional dialogue systems, which often generate mundane and generic responses like 'Okay, I am fine.' based solely on the dialogue context, knowledge-based systems integrate dialogue history with relevant knowledge to produce more informative and content-rich responses. This advancement signifies a major leap in the development of dialogue systems, moving from basic interaction to engaging, knowledge-driven exchanges. However, despite the integration of dialogue history and knowledge by previous knowledge dialogue systems \cite{wu2022improving,chen2021unsupervised,zheng2021knowledge} through various methodologies, they overlooked a crucial issue: ensuring the timeliness of knowledge. These systems were limited to the knowledge available in their training corpora, and once the model training was completed, this knowledge base ceased to update. As a result, regardless of the ingenious methods of knowledge integration proposed by these works, their practical applicability remains limited due to the static nature of their knowledge sources.

Consequently, internet-based knowledge dialogue systems have emerged \cite{zhou2022sinc,komeili2022internet}, utilizing search engines as a dynamic source of knowledge, significantly alleviating the issue of knowledge timeliness. Typically, these systems are divided into three tasks just like Figure \ref{fig1}: \emph{Retrieval Decision}, which determines if the current dialogue necessitates external knowledge retrieval—for simple contexts like greetings, external knowledge is often unnecessary; \emph{Query Generation}, which formulates queries for knowledge retrieval from search engines; and \emph{Response Generation}, which synthesizes the retrieved knowledge with the dialogue context to formulate replies. Even large-scale pre-trained language models like ChatGPT are now exploring integration with internet retrieval, underscoring the significance of these systems.

In previous research, most scholars \cite{hu2024dynamically,zhou2022sinc,komeili2022internet} have overlooked the Retrieval Decision task, assuming that all dialogues require additional knowledge to proceed. This assumption can lead to several issues, such as increased processing time, and potential information overload, where irrelevant external knowledge might overshadow the natural flow of the conversation. Furthermore, by employing separate models for Query Generation and Response Generation, these approaches introduce additional challenges. These include difficulty in maintaining coherence between the generated query and the subsequent response, and the complexity of integrating multiple models, which can increase the cost of system deployment, posing significant obstacles to the practical application of knowledge dialogue systems.

To tackle the challenges previously outlined, we present the 'UniRQR,' a Unified model for Retrieval Decision, Query, and Response Generation. UniRQR utilizes both prompt and multi-task learning methodologies to maximize the capabilities of pre-trained models \cite{shao2021cpt,lewis2020bart}. By employing a single, unified model, it adeptly manages the trio of tasks: Retrieval Decision, Query Generation, and Response Generation. The purpose of UniRQR is grounded in the intuitive understanding that the three tasks require different perspectives of dialogue context analysis. For instance, while Retrieval Decision and Query Generation tends to summarize the dialogue context, Response Generation demands a deeper comprehension of the emotional and informational flow within the conversation. This varying focus implies that these tasks should exhibit significant interactivity and synergy. To validate this hypothesis, we conducted a comprehensive set of experiments. The results from these experiments confirmed the accuracy of our hypothesis regarding the interconnectedness and synergy among the tasks within UniRQR.

Our primary contributions are as follows:
\begin{itemize}
	\item We introduce 'UniRQR,' a novel unified model for internet-based knowledge dialogue systems. Unlike previous systems that use separate models for each task and typically overlook the critical Retrieval Decision task, UniRQR efficiently handles Retrieval Decision, Query Generation, and Response Generation within a single model.
	
	\item Our model demonstrates superior performance compared to individual task baselines which indicates that integrating the three tasks into a single model not only maintains but also enhances the quality of each task. It also achieves comparable results when contrasted with SOTA systems that deploy separate, specialized models for each task.
	
	\item  Our work delves into the synergistic relationship among the tasks in Internet-based knowledge dialogue systems, offering insights into how these interactions enhance overall system performance.
\end{itemize}
\begin{figure*}[htbp]
	\centering
	\includegraphics[width=2\columnwidth]{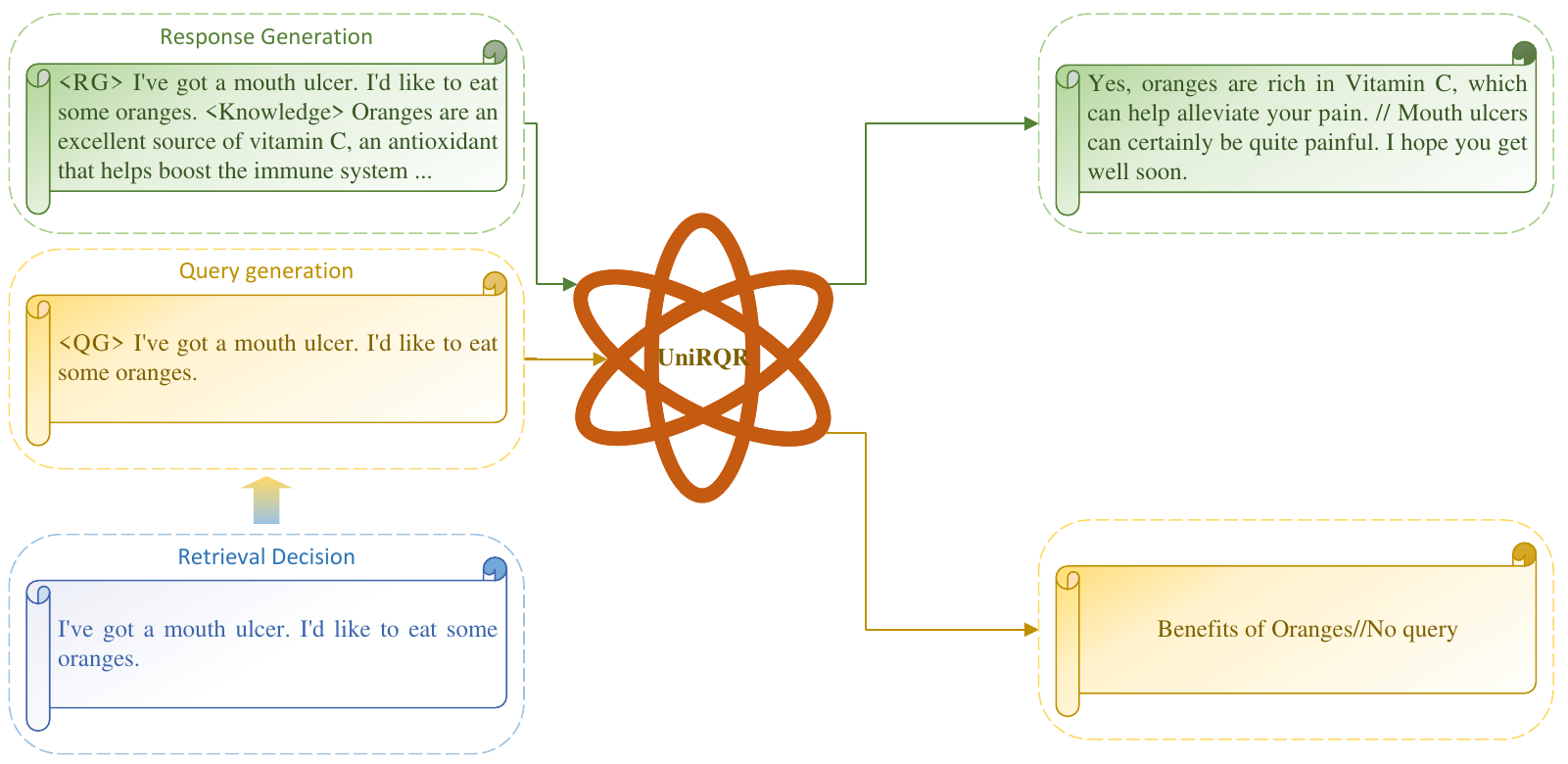}
	\caption{This is an overview diagram of the UniRQR system. First, we transform the input-output format of the Retrieval Decision task to align with the Query Generation task. Then, we use different prompts to either complete the Response Generation task or the combined RD/QG tasks.}
	\label{fig2}
\end{figure*}
\section{Related Work}
With the advent of pre-trained language models \cite{shao2021cpt,lewis2020bart}, open-domain dialogue systems \cite{roller2021recipes,kann2022open,ji2022achieving} have garnered increased attention. Unlike task-oriented dialogue systems, which are limited to specific topics, open-domain dialogues offer users the freedom to converse on any subject. However, inherent limitations in the training of Natural Language Generation (NLG) tasks \cite{xu2021adaptive} have led to a common issue with general open-domain dialogue models: they tend to generate responses that are often boring and lack meaningful content. 

To address the limitations of general open-domain dialogue models, researchers have explored the integration of external information sources with dialogue context \cite{varshney2023emotion,li2022knowledge,prabhumoye2021focused}, aiming to produce more meaningful responses. Among these approaches, leveraging knowledge as an external information source has been the most prevalent. This led to the development of knowledge-based dialogue systems, where researchers employ a variety of techniques to infuse external knowledge into dialogue contexts \cite{wu2022improving,lin2020generating}, enhancing the significance of the model-generated responses. However, previous studies have commonly neglected the aspect of knowledge timeliness. These systems were constrained to the knowledge present in their training datasets, becoming outdated post-training. Therefore, despite innovative approaches to knowledge integration, the practical utility of these systems is compromised by their reliance on static knowledge sources. 

Internet-based knowledge dialogue systems have thus arisen \cite{zhou2022sinc,komeili2022internet}, using search engines to dynamically source knowledge and address the issue of timeliness. These systems typically encompass three main tasks: Retrieval Decision, assessing the need for external knowledge in a dialogue; Query Generation, crafting search queries; and Response Generation, integrating retrieved knowledge into dialogue replies. Previous studies, like \cite{hu2024dynamically,zhou2022sinc} and \cite{komeili2022internet}, all neglected the Retrieval Decision task. They operated under the assumption that dialogues universally require additional knowledge, leading to increased processing time, and potential information overload, where irrelevant external knowledge might overshadow the natural flow of the conversation. \cite{shuster2022language} pre-trained their models, which substantially increased training costs. Moreover, \cite{hu2024dynamically,zhou2022sinc,komeili2022internet} developed distinct models for Query Generation and Response Generation. This approach posed challenges in maintaining coherence between these stages and resulted in higher system deployment costs, thereby hindering the practical application of knowledge dialogue systems.

\section{Methodology}
\subsection{Task Formalization}
In our framework, we consider a sample $D = \{ ({U},{Y})\} $ which consists of two components. The dialogue history is denoted as ${U} = (u_1,u_2,...,u_{n_u})$, encapsulating the sequence of utterances in a conversation. During the Retrieval Decision stage, if it is determined that additional knowledge is required for an improved response generation, the Query Generation module is invoked. This module processes ${U}$ to generate a query ${Q}=(q_1,q_2,...,q_{n_q})$. Subsequently, ${Q}$ is used to retrieve relevant knowledge ${K}=(k_1,k_2,...,k_{n_k})$ from a search engine. Finally, in the Response Generation stage, the dialogue history ${U}$ and the retrieved knowledge ${K}$ are integrated to generate the response ${Y}=(y_1,y_2,...,y_{n_y})$. Here, and $n_u$, $n_q$, $n_k$ and $n_y$ denote the length of dialogue history, query, knowledge and response. In the training stage, ${Q}$ and ${K}$ is provided by dataset.
\subsection{Model Overview}
UniRQR is a simple but efficient model, tailored for Internet-based knowledge dialogue systems. Depending on the dataset, it utilizes CPT \cite{shao2021cpt} or BART \cite{lewis2020bart} as its backbone, seamlessly handling three core tasks: Retrieval Decision, Query Generation, and Response Generation. This design ensures the model's adaptability across various linguistic contexts, while its integrated approach streamlines the workflow and reduces computational overhead. 
\subsubsection{Retrieval Decision}
Retrieval Decision, a crucial function within UniRQR, determines if external knowledge retrieval is necessary for constructing responses, based on the current dialogue context. For some scenarios, like basic greetings, the model adeptly generates fitting responses using just the dialogue context, eliminating the need for external information. Prior systems typically bypassed this evaluation, operating under the assumption that all conversations required supplemental knowledge. This approach often resulted in unnecessary complexity and a propensity for integrating irrelevant information, thereby diluting the conversational relevance and efficiency.
\subsubsection{Query Generation}
The Query Generation component of UniRQR can formulate precise queries for knowledge retrieval from search engine. This process is initiated only when the Retrieval Decision task deems it necessary. The model intelligently condenses and abstracts the key elements of the dialogue context into a query, aimed at fetching the most relevant external knowledge from search engines. This capability ensures that the retrieved information is precisely tailored to augment the dialogue, enhancing the depth and accuracy of the system’s responses.
\subsubsection{Response Generation}
In the Response Generation phase of UniRQR, the model's approach adapts based on the availability of external knowledge. If search queries have retrieved relevant external knowledge, the model integrates this with the dialogue context, creating responses that are contextually pertinent and enriched with this knowledge. In cases where Query Generation is not invoked or external knowledge is not retrieved, UniRQR skillfully generates responses based solely on the existing dialogue context.

The overview of UniRQR is shown in Figure \ref{fig2}.

\subsection{Prompt Engineering}
 We introduce task-specific prompts into the respective inputs. This approach is designed to refine the model's focus on each particular task, enhancing its ability to execute that task effectively. Specifically, we have developed three varieties of prompts: special token, discrete and continuous, each tailored to encapsulate task-specific information.
\begin{itemize}
\item Special Token: Utilizing special tokens to differentiate between tasks offers a straightforward and effective approach. Specifically, for the Retrieval Decision and Query Generation tasks, we prepend the input with a unique special token, $<QG>$. Similarly, for the Response Generation task, the input is augmented with a different special token, $<RG>$. This method provides a clear and simple way for the model to distinguish between the tasks, enhancing task-specific processing efficiency.
\item Discrete Prompts: Drawing from the significant achievements seen in the use of discrete prompts across various natural language processing endeavors \cite{sanh2022multitask,liu2023disentangled}, we have crafted specialized cloze templates. These templates are meticulously designed to align with the unique attributes of each distinct task in our system.
\item Continuous Prompts: Leveraging advancements in continuous prompt learning \cite{zhou2022dual,ju2023continuous}, we aim to minimize the need for manual annotation of templates typically required for obtaining discrete prompts. In our approach, a prompt is conceptualized as a trainable dense vector within a continuous space, contrasting with the natural language text instructions characteristic of discrete prompts.
\end{itemize}
Typically, when multi-tasking with prompts, there's an observable trade-off where the performance of each task may slightly diminish. However, in the case of UniRQR, the tasks of Retrieval Decision, Query Generation, and Response Generation interact in a mutually beneficial manner. This synergy, in fact, enables UniRQR to surpass the performance of baseline models. We will discuss the phenomenon in Section \ref{sec5}.
\subsection{Context Representation}
Evidently, Retrieval Decision inherently differs from the others as it is a classification task. To align it with the generative nature of the other tasks, we transformed the label format of the Retrieval Decision task into a generative format, consistent with the Query Generation task. Specifically, if the model determines that retrieval is unnecessary, it generates the phrase "No Query", Conversely, if retrieval is deemed necessary, the model proceeds to generate a specific query. This strategy results in two distinct input formats within our system, one tailored for query-related tasks,
\begin{equation}
	X_q = (prompt_q;Context)
\end{equation}
and the other for response generation tasks.
\begin{equation}
	X_r = (prompt_r;Context;[SEP];Knowledge)
\end{equation}
\begin{equation}
	Context = (user: utterance_1; bot: utterance_2;...)
\end{equation}
Here, the $Context$ is a concatenation of the dialogue history, explicitly augmented with the Speaker information. Meanwhile, the $Knowledge$ consists of the information retrieved. Should the current dialogue not necessitate external knowledge, the knowledge segment remains empty, allowing the model to adapt its response generation based on the dialogue history alone. 
\subsection{Training Strategy}
Given that our UniRQR model is adept at simultaneously executing these three tasks, it adapts its output based on the nature of the input. The output generation mechanism of UniRQR can be described as follows:
\begin{itemize}
\item For Retrieval Decision and Query Generation: When presented with input related to these tasks, UniRQR produces outputs that either determine the necessity of information retrieval or formulate a specific query for external knowledge.
\begin{equation}
	Y^{'}_q = UniRQR(X_q|W_\theta)
\end{equation}
\item For Response Generation: In this case, UniRQR, upon receiving dialogue context and, if applicable, retrieved knowledge, generates a coherent and contextually relevant response.
\begin{equation}
	Y^{'}_r = UniRQR(X_r|W_\theta)
\end{equation}
\end{itemize}
where $\theta$ denotes the model parameters, and the loss items corresponding to the three tasks in UniRQR are defined as follows.
\begin{equation}
	{L_{rd}}(\theta) =  - \frac{1}{{\left| N_{rd} \right|}}\sum\limits_{t = 1}^{\left| N_{rd} \right|}CrossEntropy(Y_q|Y^{'}_q)
\end{equation}
\begin{equation}
	{L_{qg}}(\theta) =  - \frac{1}{{\left| N_{qg} \right|}}\sum\limits_{t = 1}^{\left| N_{qg} \right|}CrossEntropy(Y_q|Y^{'}_q)
\end{equation}
\begin{equation}
	{L_{rg}}(\theta) =  - \frac{1}{{\left| N_{rg} \right|}}\sum\limits_{t = 1}^{\left| N_{rg} \right|}CrossEntropy(Y_r|Y^{'}_r)
\end{equation}
where $N_{rd}$, $N_{qg}$ and $N_{rg}$ represent the total number of instances in Retrieval Decision, Query Generation, Response Generation.
 
Our model is trained using a multi-task learning approach, thus allowing it to simultaneously optimize for all three tasks. The model's parameters are refined through end-to-end optimization using the cross-entropy loss as the objective function.
\begin{equation}
	\mathcal{L(\theta)} = \alpha{L_{rd}}(\theta)+\beta{L_{qg}}(\theta)+\gamma{L_{rg}}(\theta)
\end{equation}
Here, $\alpha$, $\beta$, $\gamma$ are hyperparameters.
\begin{table*}[!t]
	\centering
	\resizebox{\linewidth}{!}{
		\begin{tabular}{c|ccc|ccc|ccc}
			\toprule
			\multirow{2}{*}{\textbf{Models}} & \multicolumn{3}{c|}{\textbf{Dusinc}} & \multicolumn{3}{c|}{\textbf{WizInt-Single}} & \multicolumn{3}{c}{\textbf{WizInt-Multi}}\\
			& \textbf{F1}   & \textbf{BLEU-1}  & \textbf{BLEU-2}  &\textbf{F1}   & \textbf{BLEU-1}  & \textbf{BLEU2}  & \textbf{F1}   & \textbf{BLEU-1}  & \textbf{BLEU-2} \\ \hline
			Transformer\cite{zhou2022sinc}              & 16.3\%   & 10.8\%  & 10.6\%  & \multicolumn{3}{c|}{\emph{No report}}  & \multicolumn{3}{c}{\emph{No report}}         \\ 
			Backbone                       & 42.4\%   & 35.1\%  &31.7\%  &  27.4\%       & 24.8\%      & 16.2\%     & 41.4\%       & 45.3\%   &30.0\%        \\ 
			PDAML\cite{zeng2023personalized}                       &44.6\%         & \underline{38.8\%}       & 33.4\%       & 26.9\%           & 26.6\%      & 18.3\%    & 42.7\%       &47.1\%       &32.6\%           \\ 
			DRKQG\cite{hu2024dynamically}         &\textbf{47.1\%}   & \textbf{41.6\%} &\textbf{37.0\%}    &  \underline{27.4\%}       & 24.8\%      & 16.2\%       & \textbf{44.0\%}  & \textbf{49.8\%}  &\textbf{33.9\%}        \\  
			UniRQR(w RD)         &40.9\%   & 37.0\% & 32.0\%     & 25.3\%          & \textbf{31.5\%}              & \textbf{22.2\%}        & 35.3\%  & 45.3\%  &\underline{33.8\%}        \\ 
			\textbf{UniRQR(w/o RD)}        &\underline{46.6\%}   & 38.5\% &\underline{34.6\%}     & \textbf{27.5\%}       &  \underline{30.4\%}      & \underline{18.4\%}       & \underline{43.5\%}  & \underline{47.6\%} & 30.9\% \\
			\bottomrule
	\end{tabular}}
	\caption{The table presents the experimental results for query generation. The 'w RD' row indicates the performance when the retrieval decision task is considered in the query generation, while 'w/o RD' signifies scenarios where the retrieval decision task is not factored in. Due to the nature of the evaluation metrics used in previous works, which particularly emphasize instances requiring retrieval, the performance under the 'w/o RD' condition warrants special attention for a fair comparison.}
	\label{tab1}
\end{table*}
\begin{table*}[!t]
	\centering
	\begin{tabular*}{\hsize}{@{}@{\extracolsep{\fill}}cccccc@{}}
		\toprule
		\textbf{Model}      & \textbf{F1} & \textbf{BLEU-1} & \textbf{BLEU-2} & \textbf{DISTINCT-1} & \textbf{DISTINCT-2}\\ \midrule
		Transformer\cite{zhou2022sinc} & 20.0\%            & 13.7\%  & 8.8\%          & 14.8\% & 53.5\%          \\
		CPT-Large\cite{shao2021cpt}         & 29.5\%          & 24.6\%  & 17.1\%          & \textbf{17.3\%} & 66.0\%          \\
		KIC\cite{lin2020generating}         & 28.3\%          & 22.3\%  & 14.3\%          & 7.5\%  &39.2\%          \\
		CPT-sw\cite{cao2020pretrained}              & 30.3\%          & 23.7\%   &16.3\%          & 16.4\%  &64.1\%          \\
		DoHA\cite{prabhumoye2021focused}            & 31.4\%          & 25.5\%  &18.4\%          & 16.8\%   &\textbf{67.1\%}           \\\midrule
		DRKQG\cite{hu2024dynamically}       & 33.3\%        & 30.8\%    & 21.6\%     &  11.7\% & 58.8\%          				\\ 
		\textbf{UniRQR}       & \textbf{34.0\%}          &\textbf{32.3\%}    & \textbf{22.8\%}     &  14.7\% & 68.0\%          \\ \bottomrule
	\end{tabular*}
	\caption{The table above details the experimental results for response generation on Dusinc. The performance of UniRQR surpasses that of most baseline models, with its F1 and BLEU scores nearing the current state-of-the-art (SOTA) models. Additionally, UniRQR's Distinct scores are comparable to the majority of these baseline models.}
	\label{tab2}
\end{table*}
\begin{table*}[!t]
	\centering
	\begin{tabular*}{\hsize}{@{}@{\extracolsep{\fill}}c|cccc|cccc@{}}
		\toprule
		\multirow{2}{*}{\textbf{Models}} & \multicolumn{4}{c|}{\textbf{WizInt-Single}}                                                                                          & \multicolumn{4}{c}{\textbf{WizInt-Multi}}                                                                  \\
		& \textbf{F1} & \multicolumn{1}{l}{\textbf{KF1}} & \multicolumn{1}{l}{\textbf{BLEU-1}} & \textbf{D-BLEU-2} & \textbf{F1} & \textbf{KF1} & \textbf{BLEU-1} & \multicolumn{1}{l}{\textbf{BLEU-2}} \\ \hline
		Bart-Large*\cite{lewis2020bart}                 & 18.4\%          &21.8\%     & 20.1\%         & 5.73\%            &20.3\%          & 20.2\%        & 22.2\%              & 6.72\%       \\
		Bart-sw\cite{cao2020pretrained}        & 19.1\%          & 21.5\%      & 19.9\%         & 5.81\%                   & 20.1\%               & 21.2\%             & 20.3\%                   & 5.99\%          \\
		DoHA\cite{prabhumoye2021focused}             & 18.6\%          & 22.8\%       & 19.8\%         & 5.64\%             & 19.6\%          & 22.3\%        & 21.1\%              & 6.31\%    \\
		Bart-Large\cite{lewis2020bart}                & \textbf{25.4\%}          & 23.1\%       & \multicolumn{2}{c|}{\multirow{3}{*}{\emph{No report}}}            & \multicolumn{4}{c}{\multirow{3}{*}{\emph{No report}}}             \\
		BlenderBot-400M                 & 22.0\%         & 22.8\%             &   &      &       &    &                  &          \\
		BlenderBot-2.7B                  & 21.7\%          & 23.3\%            &  &        &              &              &                   &             \\ \hline
		SeeKeR\cite{shuster2022language}                          & 24.5\%          & 21.6\%             & \multicolumn{2}{c|}{\emph{No report}}                 & \multicolumn{4}{c|}{\emph{No report}}        \\
		DRKQG\cite{hu2024dynamically}            & 19.5\%          & 23.3\%      & 20.8\%        & 6.08\%         & 20.3\%          &22.7\%         & 21.2\%              & 6.30\%     \\
		\textbf{UniRQR}            & 23.1\%          & \textbf{28.3\%}      & \textbf{24.6\%}        & \textbf{7.91\%}        & \textbf{26.1\%}          & \textbf{26.3\%}      & \textbf{25.0\%}        & \textbf{7.76\%}      \\
		\bottomrule
	\end{tabular*}
	\caption{The table presented above shows the experimental results for response generation on the WizInt dataset. UniRQR achieved state-of-the-art (SOTA) results across most metrics. Critically, the Bart-large model is listed twice in the table. This duplication stems from the fact that various researchers have reported differing results for Bart-large on the WizInt dataset in previous works. To ensure a comprehensive and balanced comparison, we have chosen to include both sets of results for Bart-large.}
	\label{tab3}
\end{table*}
\section{Experiments}
The implementation details of UniRQR will be thoroughly presented in Appendix A.
\subsection{Dataset}
Our evaluation was conducted on both the \textbf{WizInt} \cite{komeili2022internet} and \textbf{Dusinc} \cite{zhou2022sinc} datasets, catering to English and Chinese dialogues respectively. For WizInt, following the approach of \cite{hu2024dynamically} and \cite{komeili2022internet}, we utilized two data processing methods: single-turn dialogue samples (WizInt-Single) and multi-turn dialogue samples (WizInt-Multi). WizInt-Single focuses on extracting individual dialogue turns for evaluating single-turn interaction efficiency. In contrast, WizInt-Multi covers comprehensive multi-turn dialogues, offering a more intricate and realistic testing environment.

Meanwhile, the DuSinc dataset, featuring diverse open-domain Chinese dialogues, was used for complementary evaluation. It comprises 2,200 dialogues with over 11,466 turns, annotated with queries and knowledge pertinent to each dialogue, thus enabling both training and validation of our model across a spectrum of conversational complexities.
\subsection{Comparison Models and Metrics}
\begin{table*}[!t]
	\centering
	\resizebox{\linewidth}{!}{
		\begin{tabular}{c|cccccc|cccccc|cccccc}
			\toprule
			\multirow{2}{*}{\textbf{Model}} & \multicolumn{6}{c|}{\textbf{Dusinc}}   & \multicolumn{6}{c|}{\textbf{WizInt-Single}}                   & \multicolumn{6}{c}{\textbf{WizInt-Multi}}                \\
			& \textbf{Acc}& \textbf{TPR}& \textbf{TNR} & \textbf{F1} & \textbf{BLEU-1} & \textbf{BLEU-2} & \textbf{Acc}& \textbf{TPR}& \textbf{TNR}	& \textbf{F1} & \textbf{BLEU-1} & \textbf{BLEU-2} & \textbf{Acc}& \textbf{TPR}& \textbf{TNR} & \textbf{F1} & \textbf{BLEU-1} & \textbf{BLEU-2} \\ \midrule
			UniRQR                    &68.1\%     & 80.8\%          & 50.1\%              & 46.9\%        &53.8\%      & 44.6\%          & 65.7\%              & 74.0\%  &33.6\%  & 25.3\%          & 31.5\%              & 22.2\%   & 66.0\%  &73.5\%  & 36.6\%          & 35.3\%              & 45.3\%      & 33.8\%   \\
			w/o RD           &--    &--  &--    & 55.7\%          &  62.2\%              & 50.6\%      &--    &--  &--       & 27.5\%          & 30.4\%              & 18.4\%   &--   &--    &--  & 43.5\%          & 47.6\%              & 30.9\%         \\
			w/o RG        &63.9\%   & 69.6\%              & 55.8\%           &41.4\%   & 50.7\%              & 42.4\%         & 64.3\%      & 72.2\%       & 33.6\%              & 25.3\%                  & 30.8\%    &21.4\%   & 61.1\%              & 65.1\%                  & 45.8\%    & 32.3\%              & 44.3\%                  & 36.3\%          
			\\
			w/o Knowledge     &69.7\%     & 85.7\%              & 47.9\%                  & 49.7\%        &  57.9\%      & 48.1\%             &68.6\%     & 77.6\%              & 33.5\%                  & 26.4\%        &  32.3\%      & 22.3\%    &67.9\%     & 75.8\%              & 36.6\%                  & 36.4\%        &  45.6\%      & 34.7\%                     
			\\
			w/o Knowledge \& RD      &--    &--  &--      & 56.5\%             & 64.1\%                  & 53.2\%         &--    &--  &--       & 28.5\%              & 31.6\%                  & 19.2\%      &--    &--  &--    & 44.1\%              & 46.3\%                  & 32.7\%                         
			\\ \bottomrule
	\end{tabular}}
	\caption{The table displays the results of UniRQR's ablation experiments for the Query Generation task, where we sequentially removed the Retrieval Decision and Response Generation tasks to assess their impact on QG. In these experiments, samples requiring retrieval were treated as positive samples, with TPR representing the True Positive Rate and TNR as the True Negative Rate. Due to the unavailability of the test set and the official evaluation metrics not including TPR and TNR for the Dusinc dataset, we opted to use the validation set for these assessments. Additionally, we removed knowledge from the input of the RG task to test our hypothesis about the synergistic effects of these tasks on each other.}
	\label{tab4}
\end{table*}
\begin{table*}[!t]
	\centering
	\resizebox{\linewidth}{!}{
		\begin{tabular}{c|ccccc|cccc|cccc}
			\toprule
			\multirow{2}{*}{\textbf{Model}} & \multicolumn{5}{c|}{\textbf{Dusinc}}   & \multicolumn{4}{c|}{\textbf{WizInt-Single}}                   & \multicolumn{4}{c}{\textbf{WizInt-Multi}}                \\
			& \textbf{F1} & \textbf{BLEU-1} & \textbf{BLEU-2} & \textbf{DISTINCT-1}	& \textbf{DISTINCT-2} & \textbf{F1}& \textbf{KF1}& \textbf{BLEU-1} & \textbf{BLEU-2}  & \textbf{F1}& \textbf{KF1}& \textbf{BLEU-1} & \textbf{BLEU-2} \\ \midrule
			UniRQR                    & 34.0\%          &32.3\%    & 22.8\%     &  14.7\%           & 68.0\%     & 23.1\%          & 28.3\%      & 24.6\%       & 7.91\%        & 26.1\%          & 26.3\%      & 25.0\%       & 7.76\%           \\
			w/o RD           &33.0\%     & 29.9\%          & 21.2\%              & 15.4\%        &69.0\%      & 22.3\%          & 23.5\%              & 24.8\%  & 7.7\%            & 21.8\%              & 26.0\%    &24.7\% & 7.7\%           \\
			w/o QG \& RD      &29.5\%     & 24.6\%          & 17.1\%              & 17.3\%        &66.0\%      & 22.4\%          & 27.0\%              & 23.2\%  & 7.2\%            & 24.9\%              & 26.5\%    &24.0\% & 7.4\%             
			\\ \bottomrule
	\end{tabular}}
	\caption{The ablation study for Response Generation task}
	\label{tab5}
\end{table*}
Most existing models struggle to concurrently manage the three tasks, with the Retrieval Decision task presenting a notable challenge that many previous studies have not adequately addressed, making direct comparisons for this task with existing models challenging. Thus, To provide a focused evaluation, following \cite{komeili2022internet,zhou2022sinc}, we developed separate comparison models for the query and response generation tasks. The distinct impact and challenges posed by the Retrieval Decision task are discussed in detail in Section \ref{sec5.3}. The performances of UniRQR, alongside its comparative models in query generation and response generation, are comprehensively summarized in Table \ref{tab1} and Tables \ref{tab2}, \ref{tab3}.

\subsection{Results}
Previous studies have not adequately addressed the Retrieval Decision task. Therefore, we primarily focus on the Query Generation and Response Generation tasks here. The performance of the Retrieval Decision task and its consequent impact will be discussed in detail in Section \ref{sec5.3}.

The automatic evaluation results for the Query Generation task are detailed in Table \ref{tab1}. Our methodology for calculating evaluation metrics mirrors that of \cite{zhou2022sinc}, specifically in their approach to the Query Generation task. In their method, metrics such as F1 score are computed using only those instances that necessitate knowledge retrieval. This means that if an instance requiring retrieval incorrectly generates a 'No Query' response, it significantly impacts the evaluation metrics, highlighting a model's failure in accurately identifying retrieval needs. Conversely, instances correctly identified as 'No Query' do not influence the overall metrics. This approach inherently results in a substantial decrease in performance metrics for the Query Generation task when instances of 'No Query' are considered.

Therefore, in Table \ref{tab1}, we focus on the performance of our models without the integration of the Retrieval Decision task, and our model exhibits a marked improvement in the quality of query generation compared to the backbone model. Specifically, in the context of the Query Generation task, when the Retrieval Decision is not considered, UniRQR demonstrates significant improvements over the Backbone model. On the Dusinc dataset, UniRQR shows increases of 9.9\% in F1 score, 9.7\% in BLEU-1, and 9.1\% in BLEU-2. Similarly, in the WizInt-Multi dataset, there are improvements of 5\%, 4.9\%, and 3\% in F1, BLEU-1, and BLEU-2 scores respectively. This improvement is significant, especially when considering that our model remains competitive even against specialized systems like DRKQG, which designed a explicit model for the QG task.

The automatic evaluation results for the Response Generation task are comprehensively presented in Tables \ref{tab2} and \ref{tab3}. As indicated by these results, UniRQR consistently surpasses the backbone model across all datasets on most metrics. Specifically,in the Dusinc dataset, UniRQR achieved increases of 15.2\% in F1 score, 30.1\% in BLEU-1, and 33.3\% in BLEU-2. In the WizInt-Single dataset, the improvements were even more remarkable: 25.6\% in F1, 29.8\% in KF1, 22.4\% in BLEU-1, and 38.0\% in BLEU-2. Additionally, in the WizInt-Multi dataset, UniRQR showed enhancements of 28.6\% in F1, 30.2\% in KF1, 17.1\% in BLEU-1, and 15.4\% in BLEU-2. 

The experimental results conclusively demonstrate that in Internet-based knowledge dialogue systems, the three tasks of Retrieval Decision, Query Generation, and Response Generation can mutually enhance each other's performance. UniRQR, utilizing just a single model, is able to effectively execute these three tasks, achieving and in some cases surpassing the performance of systems that employ separate, specialized models for each task, such as \cite{hu2024dynamically,zhou2022sinc}. Additionally, unlike the Seeker \cite{shuster2022language} which requires extra pre-training, UniRQR capitalizes on the inherent capabilities of pre-trained language models simply through appropriate training strategies.
\section{Analysis and Discussion}\label{sec5}
\subsection{Ablation Study}
Ablation studies were conducted for the Query Generation (QG) and Response Generation (RG) tasks, with the findings detailed in Tables \ref{tab4} and \ref{tab5}. For the QG task, it is evident that excluding the RG task results in a performance decline, regardless of whether the Retrieval Decision (RD) task is considered or not. Similarly, for the RG task, disregarding either the RD or QG tasks also leads to a decrease in performance. This pattern observed in the experiments substantiates our hypothesis that the three tasks are mutually enhancing. The decline in performance when any of these tasks are removed highlights their synergistic relationship, where each task significantly contributes to and benefits from the integrated functioning of the others.
\subsection{Unified Input Synergy}
As previously discussed, we believe the mutual enhancement among the three tasks stems from their distinct angles in interpreting dialogue history, leading to a more comprehensive utilization of contextual information. Following this line of thought, we hypothesized that removing the knowledge component from the RG task, thus standardizing the input format across all three tasks to include only the dialogue history and speaker information, might yield greater benefits for the RD and QG tasks. The results from our experiments, as shown in Table \ref{tab4}, validate this hypothesis. When knowledge is not included as an input for the RG task, both the QG and RD tasks demonstrate improved performance. This outcome supports our notion that a uniform input format across these tasks enables a more effective and cohesive use of context, thereby enhancing the overall performance of each individual task.
\subsection{Effect of RD Task}\label{sec5.3}
\begin{table}[!t]
	\centering
	\resizebox{\linewidth}{!}{
		\begin{tabular}{c|cc|cc|cc}
			\toprule
			\multirow{2}{*}{\textbf{Model}} & \multicolumn{2}{c|}{\textbf{Dusinc}}   & \multicolumn{2}{c|}{\textbf{WizInt-Single}}                   & \multicolumn{2}{c}{\textbf{WizInt-Multi}}                \\
		& \textbf{TPR}& \textbf{TNR}& \textbf{TPR}& \textbf{TNR}& \textbf{TPR}& \textbf{TNR} \\ \midrule
			UniRQR            & 80.8\%          & 50.1\%                     & 74.0\%  &33.6\%    &73.5\%  & 36.6\%      \\
			Backbone           &71.1\%    &63.3\%     &95.6\%  &12.1\%    &86.7\%    &14.9\%      \\
		 \bottomrule
	\end{tabular}}
	\caption{The table displays the results }
	\label{tab6}
\end{table}
We conducted a detailed statistical analysis of the performance of the RD task. UniRQR demonstrated commendable predictive reliability in determining the necessity of retrieval. This contrasts with previous works, which often defaulted to assuming retrieval was always necessary, resulting in a zero probability of true negatives. 

We implemented backbone models for the Retrieval Decision task, and compared them with UniRQR on the Dusinc and WizInt datasets, as shown in Table \ref{tab6}. The results indicate that, for the Dusinc dataset, the backbone model achieved a more balanced classification: it showed similar accuracy rates for predicting both retrieval-required instances (True Positive Rate) and non-retrieval instances (True Negative Rate), whereas UniRQR exhibited a higher accuracy in predicting retrieval-required instances. However, in the case of the WizInt dataset, which has a highly imbalanced sample distribution with 75\% of the samples requiring retrieval, the backbone model tended to predict retrieval-required for most instances. In fact, in 10 random tests conducted with different seeds, in 8 of these tests, the model predicted all the test samples as requiring retrieval. This observation indirectly illustrates that integrating the Query Generation and Response Generation tasks can significantly enhance the reliability of the Retrieval Decision task, preventing it from falling into certain biases, especially in scenarios with imbalanced sample distributions. 

\subsection{Effect of Prompt}
The impact of various prompts on UniRQR's performance is detailed in Appendix B. Our experiments revealed that different prompts had limited influence on UniRQR's effectiveness. Therefore, for simplicity and convenience in our model, we consistently used special tokens as prompts in the reported results above. We posit that in UniRQR, the role of a prompt leans more towards acting as an identifier, crucial for discerning the current task at hand. This approach streamlines the model's functionality, ensuring that it efficiently recognizes and responds to the task-specific requirements without the need for complex or varied prompt structures.

\subsection{Future Work}
As we foresee internet-based knowledge dialogue systems becoming an increasingly prominent research focus, we propose that future efforts should concentrate on further refining models to enhance the synergistic effects observed in our study. Additionally, we encourage future researchers to explore these synergistic interactions on a larger scale, particularly by utilizing large pre-trained language models. Such investigations could not only amplify the emergent capabilities of these models but also potentially offer a degree of interpretability to their emergent abilities. By delving into the dynamics of these advanced models, future research could unlock deeper insights into the complexities of dialogue systems and their interactions with vast knowledge bases, paving the way for more sophisticated and intuitive AI-driven communication tools.

\section{Conclusion}
In summary, our research demonstrates that UniRQR, an innovative model for internet-based knowledge dialogue systems, effectively handles the integration of Retrieval Decision, Query Generation, and Response Generation tasks within a single framework. Our experimental results reveal that these tasks, when synergistically combined, can significantly enhance each other, leading to a performance that matches or even exceeds systems with separate models for each task. UniRQR's efficiency is further highlighted by its ability to leverage pre-trained language models through strategic use of prompts and training methods, negating the need for additional complex pre-training. This study not only showcases the potential of UniRQR in advancing dialogue systems but also contributes to the broader understanding of task integration in AI-driven communication technologies.
\appendix

\section{Implement Details}
We implemented our model using the Pytorch framework. The learning rate was set at 2e-5. For the Dusinc dataset, we set $\aleph$ to 0.2, and both $\beta$ and $\gamma$ to 1. For the WizInt dataset, $\aleph$ was set 1.2. Our training batch size was 8, and all training was completed on a single RTX 3090.
\begin{table*}[htbp]
	\centering
	\resizebox{\linewidth}{!}{
		\begin{tabular}{r|ccc|ccc|ccc}
			\toprule
			\multirow{2}{*}{\textbf{Models}} & \multicolumn{3}{c|}{\textbf{Dusinc}} & \multicolumn{3}{c|}{\textbf{WizInt-Single}} & \multicolumn{3}{c}{\textbf{WizInt-Multi}}\\
			& \textbf{F1}   & \textbf{BLEU-1}  & \textbf{BLEU-2}  &\textbf{F1}   & \textbf{BLEU-1}  & \textbf{BLEU-2}  & \textbf{F1}   & \textbf{BLEU-1}  & \textbf{BLEU-2} \\ \hline
			\textbf{UniRQR(w RD)}         &   &  &      &          &              &         &   &   &        \\ 
			-Discrete        &+2.4\%   & +3.3\% &+2.1\%     & +3.1\%       & +4.6\%      & -1.5\%       & +2.3\%  & +3.8\% & -1.6\% \\
			-Continuous         &+2.9\%   & +2.0\% &+1.6\%     & +3.4\%       & +3.7\%      & +1.0\%       & +2.5\%  & +2.2\% & +1.8\% \\ \midrule
			\textbf{UniRQR(w/o RD)}         &   &  &     &           &              &         &   &  &        \\ 
			-Discrete        &+3.1\%   & +3.1\% &+4.7\%     & +2.6\%       & +3.3\%      & +0.5\%       & -0.3\%  & +4.4\% & +5.1\% \\
			-Continuous         &+6.2\%   & +5.3\% &+6.6\%     & +2.1\%       & +1.4\%      & +1.5\%       & +2.3\%  & +3.0\% & +4.2\% \\  
			\bottomrule
	\end{tabular}}
	\caption{This table displays the experimental results of different types of prompts in the QG task. We consider UniRQR, which uses special tokens as prompts, as the baseline model. The 'w RD' row indicates the performance when the retrieval decision task is considered in the query generation, while 'w/o RD' signifies scenarios where the retrieval decision task is not factored in. Due to the nature of the evaluation metrics used in previous works, which particularly emphasize instances requiring retrieval, the performance under the 'w/o RD' condition warrants special attention for a fair comparison.}
	\label{tab1}
\end{table*}
\begin{table*}[htbp]
	\centering
	\resizebox{\linewidth}{!}{
		\begin{tabular}{r|ccc|cccc|cccc}
			\toprule
			\multirow{2}{*}{\textbf{Models}} & \multicolumn{3}{c|}{\textbf{Dusinc}} & \multicolumn{4}{c|}{\textbf{WizInt-Single}} & \multicolumn{4}{c}{\textbf{WizInt-Multi}}\\
			& \textbf{F1}   & \textbf{BLEU-1}  & \textbf{BLEU-2}  &\textbf{F1} & \textbf{KF1}  & \textbf{BLEU-1}  & \textbf{BLEU-2}  & \textbf{F1} & \textbf{KF1}  & \textbf{BLEU-1}  & \textbf{BLEU-2} \\ \hline
			\textbf{UniRQR}         &   &  &      &          &              &         &   &   &     &  &  \\ 
			-Discrete        &-3.1\%   & +3.0\% &-2.9\%     & +2.2\%       & -3.6\%      & +1.7\%       & +3.2\%  & +4.8\% & +0.6\% & +3.4\% & +3.9\% \\
			-Continuous         &+4.5\%   & +5.6\% &+3.3\%     & +4.0\%       & +3.3\%      & -1.2\%       & +4.1\%  & +4.4\% & +3.9\%  & -2.4\% & +2.9\% \\
			\bottomrule
	\end{tabular}}
	\caption{This table displays the experimental results of different types of prompts in the RG task. We consider UniRQR, which uses special tokens as prompts, as the baseline model.}
	\label{tab2}
\end{table*}
\section{The Impact of Various Prompts}
For discrete prompts, we experimented with various templates. Similarly, for continuous prompts, we tried multiple prompt lengths. In this report, we only present the best results obtained. As shown in the Table \ref{tab1} and Table \ref{tab2}, it's true that different prompts have some impact on the final experimental results, but this impact is limited. Therefore, we ultimately decided to use simple special tokens as the prompt for UniRQR.

The templates we use can be represented as: \emph{Please generate a short query for this conversation: [X]} for Query Generation and \emph{Please generate a response for the bot to reply the user: [X]} for Response Generation. The prompt length was set to 10 for continuous prompt.

\bibliographystyle{named}
\bibliography{ijcai24}

\end{document}